\pdfoutput=1
\documentclass[aps,prd,amsmath,floats,floatfix, twocolumn,
superscriptaddress,nofootinbib,showpacs]{revtex4-2}

\usepackage[T1]{fontenc}
\usepackage[utf8]{inputenc}
\usepackage{lmodern}
\usepackage{verbatim}
\usepackage{widetable}

\usepackage[dvipsnames, usenames]{xcolor}
\definecolor{linkcolor}{rgb}{0.0,0.3,0.5}
\usepackage[hypertexnames=false, unicode, colorlinks=true, linkcolor=linkcolor,
citecolor=linkcolor, filecolor=linkcolor,urlcolor=linkcolor,
pdfusetitle]{hyperref}
\usepackage{orcidlink}

\usepackage[all]{hypcap}
\usepackage{graphicx}
\usepackage{xspace}
\usepackage{amssymb}
\usepackage[normalem]{ulem} 
\usepackage{bm} 
\usepackage{enumitem,amssymb}
\usepackage{subcaption}
\usepackage[justification=raggedright,singlelinecheck=false,font=small]{caption}
\usepackage{microtype}
\usepackage[english]{babel}
\usepackage{blindtext}
\graphicspath{%
  {figs/}%
}

\DeclareMathAlphabet{\mathpzc}{OT1}{pzc}{m}{it}

\newcommand{\lmn}{{\ell mn}}

\newcommand{\tstart}{t_\mathrm{start}}

\newcommand{\Mfdet}{M_\mathrm{f}^\mathrm{det}}

\newlist{todolist}{itemize}{2}
\setlist[todolist]{label=$\square$}

\begin{document}

\title{GW250114 reveals black hole horizon signatures}
\newcommand{\Perimeter}{\affiliation{Perimeter Institute for Theoretical Physics, Waterloo, ON N2L2Y5, Canada}}
\newcommand\anu{\affiliation{OzGrav-ANU, Centre for Gravitational Astrophysics, Research School of Physics and Research School of Astronomy \& Astrophysics, The Australian National University, ACT 2601, Australia}}
\newcommand\uib{\affiliation{IAC3–IEEC, Universitat de les Illes Balears, E-07122 Palma de Mallorca, Spain}}
\newcommand\Caltech{\affiliation{Burke Institute for Theoretical Physics and Theoretical Astrophysics 350-17, California Institute of Technology, Pasadena, CA 91125, USA}}

\author{Neil Lu \orcidlink{0000-0002-8861-9902}}
\email{neil.lu@anu.edu.au}
\anu

\author{Sizheng Ma\orcidlink{0000-0002-4645-453X}}
\email{sma2@perimeterinstitute.ca}
\thanks{Corresponding author}
\Perimeter

\author{Ornella J. Piccinni\orcidlink{0000-0001-5478-3950}}
\email{ornella.piccinni@uib.es}
\anu
\uib

\author{Yanbei Chen\orcidlink{0000-0002-9730-9463}}
\email{yanbei@caltech.edu}
\Caltech

\author{Ling Sun\orcidlink{0000-0001-7959-892X}}
\email{ling.sun@anu.edu.au}
\anu

\date{\today}
\begin{abstract}
The horizon of a black hole, the ``surface of no return'', is characterized by its rotation frequency $\Omega_H$ and surface gravity $\kappa$. A striking signature is that any infalling object appears to orbit at $\Omega_H$ due to frame dragging, while its emitted signals decay exponentially at a rate set by $\kappa$ as a consequence of gravitational redshift. 
Recent theoretical work predicts that the merger phase of gravitational waves from binary black hole coalescences carries direct imprints of the remnant horizon's properties, via a ``direct wave'' component that
(i) oscillates near $2\Omega_H$, reflecting the horizon's frame dragging and the dominant quadrupole nature of the gravitational radiation, and (ii) decays at an increasing rate characterized by $\kappa$, with additional screening from the black hole's potential barrier.
In this paper, we report observational evidence for the direct wave 
in GW250114, with a 90\% credible matched-filter signal-to-noise ratio of $15.8^{+0.1}_{-0.5}$ ($17.1^{+0.1}_{-0.4}$) in the LIGO Hanford (Livingston) detector. The measured properties are in full agreement with theoretical predictions. 
These findings establish a new observational channel
to directly measure frame-dragging effects in black hole ergospheres and explore (near-)horizon physics in dynamical, strong-gravity regimes.
\end{abstract}
\maketitle

\section*{Introduction}%

Black holes are among the most intriguing and extreme objects in the universe. Their event horizons, one-way boundaries in spacetime beyond which information cannot escape to distant observers, lie at the intersection of some of the deepest questions in modern physics, ranging from general relativity to quantum theory.
It is therefore desirable to measure the properties of horizons in astrophysical black holes.
Their one-way nature, however, renders direct observation challenging. 
Existing electromagnetic observations of black holes are largely indirect, 
including shadows and light rings around supermassive black holes imaged by the Event Horizon Telescope \cite{EventHorizonTelescope:2019dse}, X-ray reflection spectroscopy of the accretion disk inner edges~\cite{Brenneman:2006hw,Reynolds:2013qqa,Brenneman:2013oba}, and the launch of relativistic jets via the Blandford–Znajek mechanism~\cite{1977MNRAS.179..433B}. 
These electromagnetic signatures naturally lead to an appealing question: \emph{can black hole horizons be probed more directly using an alternative messenger?}

Over the past decade, the LIGO–Virgo–KAGRA detector network has observed gravitational waves from hundreds of compact binary mergers~\cite{aLIGO,aVirgo,KAGRA,GWTC-4_results,Capote:2024rmo,membersoftheLIGOScientific:2024elc}. As ripples of spacetime itself, gravitational waves carry direct imprints of spacetime geometry. 
The ``ringdown'' phase of a gravitational-wave signal is especially revealing: it mainly consists of a series of characteristic oscillations, known as quasinormal modes~\cite{vishveshwara1970, teukolsky1973, chandrasekhar1975, leaver1985, kokkotas1999,berti2009,berti2025}, excited as the remnant black hole settles into equilibrium.
These quasinormal modes encode detailed information about the remnant black hole's mass and spin \cite{PhysRevLett.26.331,PhysRevLett.34.905,Chrusciel:2012jk}. Measuring them from gravitational wave signals, an approach termed black hole spectroscopy \cite{berti2025}, therefore provides a powerful way to probe black holes. Significant efforts have been devoted to measuring these modes in observed gravitational-wave events \cite{Carullo:2019flw,bustillo2021,Ghosh:2021mrv,Finch:2022ynt, Ma:2023vvr,Ma:2023cwe,Wang:2023ljx,Correia:2023bfn,Capano:2021etf, Siegel:2023lxl,Chandra:2025ipu,Wang:2025rvn,Wang:2025baj}.

Quasinormal modes, however, are directly
tied to the surrounding light ring~\cite{Yang:2012he} rather than the horizon itself~\cite{Cardoso:2016rao}. 
Recent theoretical work~\cite{Oshita:2025qmn}, see also \cite{Mino:2008at,Zimmerman:2011dx}, has 
proposed a more direct probe for horizon properties. These studies show that during the merger stage --- the brief transition from the late inspiral of two black holes to the formation of the remnant ---
the orbital motion transitions from being dictated
by the binary's prior history
toward being
governed primarily by the intrinsic properties of the horizon of the newly formed remnant black hole. During this phase, a {\it direct wave} is emitted, oscillating at approximately twice the rotation frequency of the horizon, $2\Omega_H$, due to strong frame dragging in the ergosphere, and decaying at an increasing rate governed by the horizon surface gravity $\kappa$. This direct wave signal emerges near the peak of the gravitational-wave strain and co-exists with quasinormal modes. Importantly, Ref.~\cite{Oshita:2025qmn} suggests that such direct waves can already be detectable with the current LIGO–Virgo–KAGRA network, making their search in recently observed gravitational wave events both timely and compelling, which can open a uniquely direct avenue to probe ergosphere and horizon dynamics.

Among all detections to date, GW250114~\cite{KAGRA:2025oiz,LIGOScientific:2025obp, GW250114_082203_2025,LIGOScientific:2025snk} stands out as the loudest gravitational-wave signal from a binary black hole coalescence, with a network matched-filter signal-to-noise ratio of $\sim 80$.
Owing to its remarkable loudness, GW250114 provides an exceptional opportunity to explore the dynamical strong-gravity regime.  
This event has already enabled precision tests of Hawking's area law~\cite{KAGRA:2025oiz} and black hole spectroscopy of the merger remnant~\cite{dreyer2004,berti2006,LIGOScientific:2025obp}. Here, we demonstrate that GW250114 also serves as a powerful probe of the black hole horizon, allowing us to measure its two fundamental properties: the rotation frequency and the surface gravity.

\section*{Merger dynamics and horizon physics}%

We first summarize the physical picture for the merger stage of a binary system, as developed in theoretical studies \cite{Oshita:2025qmn,Mino:2008at,Zimmerman:2011dx}, and show how it encodes horizon physics. 
These studies build on a framework in which the perturbation of the remnant black hole by an infalling point particle is modeled, and then the predictions from this setup are compared against full numerical-relativity simulations of comparable-mass binaries. Despite its simplicity, the point-particle framework has proven remarkably powerful, offering physical insight into binary dynamics~\cite{Nichols:2011ih}, guiding tests of general relativity~\cite{Xin:2021zir,Seymour:2024kcd,Mark:2017dnq,Ma:2022xmp}, and informing waveform modeling~\cite{Buonanno:1998gg,McWilliams:2018ztb,Han:2011qz,Taracchini:2014zpa,Pan:2011gk, Wardell:2021fyy, Kuchler:2025hwx}. Most prominently, the widely used Effective One-Body formalism, central to binary black hole waveform modeling, is anchored in waveforms from a point particle orbiting a deformed black hole~\cite{Buonanno:1998gg}, supplemented with perturbations of the remnant~\cite{Pan:2013rra}.

\begin{figure}[!h]
        \includegraphics[width=\linewidth]{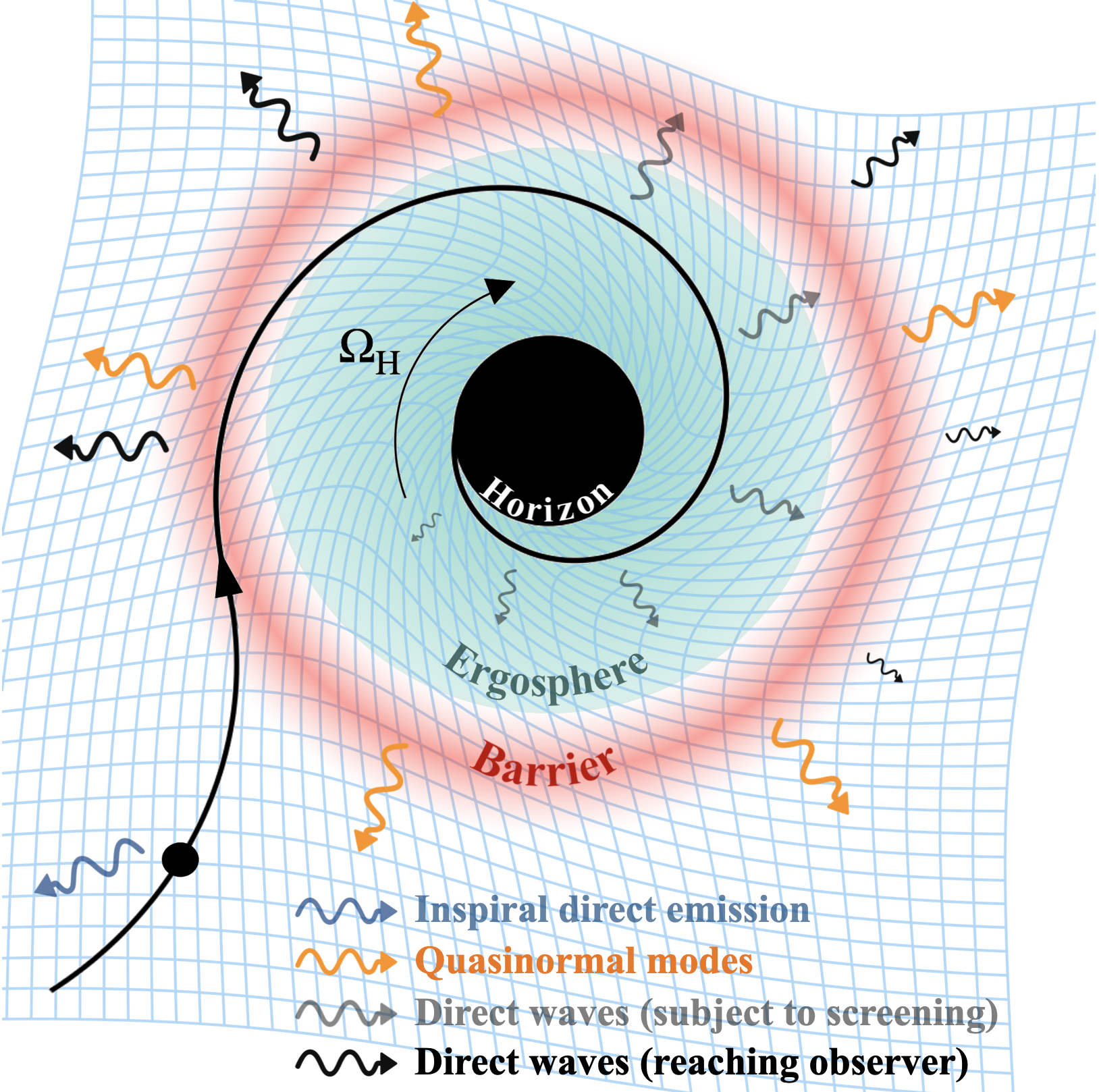}
    \caption{Sketch of the wave emission near the merger stage of a binary black hole coalescence, modeled as a point particle (small filled circle) spiraling into the remnant Kerr black hole, following the widely used Effective One-Body formalism. The red-shaded region indicates the potential barrier near the light ring, enclosing the ergosphere (green area), where the trajectory experiences strong frame dragging. Wave emission driven by the particle's motion persists from the early inspiral (blue arrow) through the final plunge, where waves (gray arrows) are gradually silenced by the remnant horizon. These waves are further screened by the potential barrier while propagating toward distant observers (black arrows). Simultaneously, quasinormal modes are excited during the barrier crossing (orange arrows). The merger portion of a gravitational-wave signal is a superposition of source-driven direct waves (black arrows)  and free quasinormal-mode oscillations (orange arrows).  
    }
    \label{fig:frame_dragging}
\end{figure}

As illustrated in Fig.~\ref{fig:frame_dragging}, we consider a point particle (small filled circle) spiraling into a Kerr black hole (central black sphere) along the curved trajectory in the equatorial plane, representing a nonprecessing binary system. The red region marks the gravitational potential barrier around the light ring, which encompasses the highly relativistic ergosphere (green area). 
During the early inspiral stage, gravitational radiation is primarily driven by orbital motion, producing waves that propagate directly to distant observers (blue arrow); this regime is well described by post-Newtonian theory~\cite{Blanchet:2013haa,poisson2014gravity}.

As the point particle passes through the black hole's potential barrier and enters the ergosphere, the characteristic free oscillations of the remnant, i.e., its quasinormal modes (orange arrows), are excited. Simultaneously, the source-driven emission continues but becomes increasingly modulated by near-horizon effects. In particular, strong frame dragging within the ergosphere governs the orbital dynamics, making it insensitive to the history of the earlier inspiral evolution. This drives the angular velocity of the infalling particle toward the horizon's intrinsic rotation frequency \cite{Mino:2008at}:
\begin{align}
    \Omega_H= \frac{\chi}{2r_+},
\end{align}
where $\chi$ is the dimensionless spin of the black hole and $r_+=1+\sqrt{1-\chi^2}$ is the radius of the outer horizon. 
As the point particle approaches the horizon, it becomes increasingly difficult for the outgoing waves (gray arrows) to escape. The near-horizon redshift effectively suppresses their amplitude, leading to an asymptotic shut-off of outgoing radiation as seen by distant observers.

The gravitational waves emitted in this regime (gray arrows) encode detailed imprints of the near-horizon dynamics. In an earlier study~\cite{Mino:2008at}, the signal was proposed to asymptote to a sinusoid with an oscillation frequency at $2\Omega_H$, and the damping timescale determined by another fundamental horizon property, the surface gravity:
\begin{align}
    \kappa=\frac{\sqrt{1-\chi^2}}{2r_+}.
\end{align}
This contribution is known as the ``horizon mode'' with a complex frequency:
\begin{align}
    \omega_H=2\Omega_H-i\kappa. \label{eq:omega_H}
\end{align}
Subsequent studies \cite{Zimmerman:2011dx,ZMC18,Oshita:2025qmn} have further shown that as these waves generated near the horizon propagate outward, they are screened by the surrounding gravitational potential barrier (red region in Fig.~\ref{fig:frame_dragging}), resulting in extra time-dependent modulations.
The gravitational waves that ultimately reach distant observers (black arrows) therefore possess evolving complex frequencies that lie \emph{close to, but not exactly at,} the horizon mode $\omega_H$. These observable signals are referred to as the ``direct waves'' \cite{Oshita:2025qmn}.\footnote{See also Refs.~\cite{deamicis2025,kuntz2025,arnaudo2025} for related discussions of other dynamical components.}

The direct waves can be interpreted as the \emph{final direct emission} from the infalling particle as it transitions from inspiral to merger, before being completely \emph{silenced} by the remnant black hole. They encode two fundamental properties of the remnant horizon: the rotation frequency $\Omega_H$ and the surface gravity $\kappa$, quantities central to the first law of black hole thermodynamics~\cite{Bardeen1973}
\begin{equation} 
   dM = \frac{\kappa }{8\pi }dA +\Omega_H dJ,
\end{equation}
where mass $M$, area $A$, and angular momentum $J$ are associated extensive variables.

The oscillation frequency of the waves reflects the frame-dragging enforced by the horizon and ergosphere, while the damping reflects the gravitational redshift set by the surface gravity. Ref.~\cite{Oshita:2025qmn} has shown that the merger portion of the signal, spanning the peak of the gravitational-wave strain, is a superposition of these source-driven direct waves and the free oscillations of quasinormal modes excited during the barrier crossing. Once the direct waves fade, the signal transitions smoothly into the final ringdown stage.

\section*{measurement of horizon properties in GW250114}%
\begin{figure}[!h]
        \includegraphics[width=\linewidth]{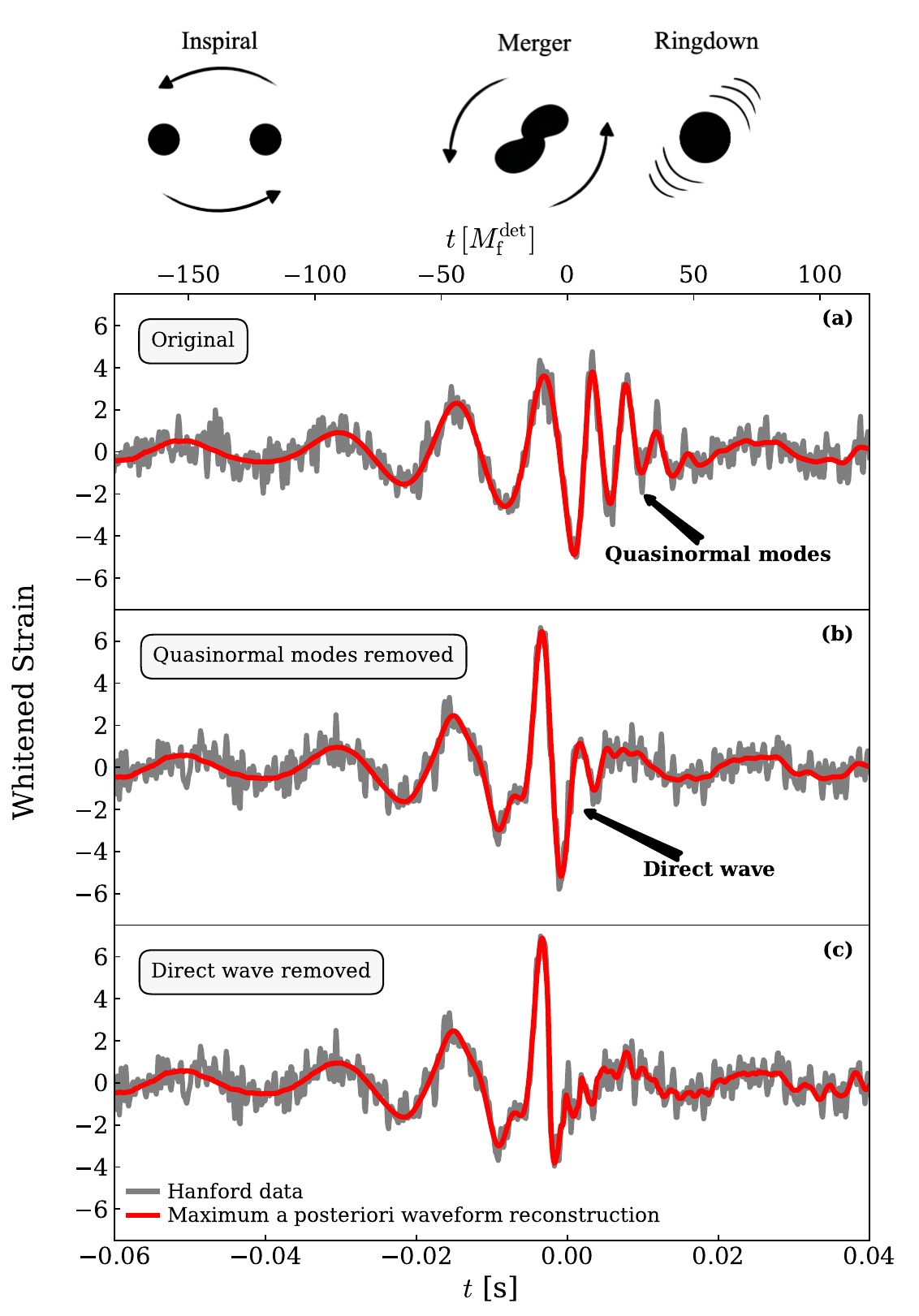}
    \caption{Whitened strain data from the GW250114 event in the LIGO Hanford detector, shown for visualization. Top: Schematic illustration of the three stages of a binary black hole coalescence: inspiral, merger, and ringdown.
    (a) Observed strain data (gray) and the corresponding maximum-a-posteriori NRSur7dq4 waveform reconstruction (red), bandpass-filtered between 20--2000~Hz.
    (b) Residual strain after removing the dominant quasinormal modes $(\ell=m=2,n=0,1,2)$. 
    (c) Residual strain after removing both the quasinormal modes and the direct wave component.
    The inferred strain peak is aligned to $t=0$.}
    \label{fig:whitened_strain}
\end{figure}

We now analyze the merger portion of GW250114. This event is from the coalescence of two black holes with component masses of $33.6^{+1.2}_{-0.8}M_\odot$ and $32.2^{+0.8}_{-1.3}M_\odot$, producing a remnant black hole with dimensionless spin $\chi_\mathrm{f} = 0.68^{+0.01}_{-0.01}$ and 
mass $M_\mathrm{f} = 62.7^{+1.0}_{-1.1}M_\odot$~\cite{KAGRA:2025oiz,LIGOScientific:2025obp, GW250114_082203_2025}.
The detector-frame remnant mass is measured to be $\Mfdet = 68.1^{+0.8}_{-0.9}M_\odot$. The whitened strain data recorded by the LIGO Hanford detector is shown in panel (a) of Fig.~\ref{fig:whitened_strain}, with the reference time defined as the inferred peak of the strain amplitude and set to $t=0$.

As discussed earlier, direct waves coexist with quasinormal modes in the merger signal. To isolate the former, we first remove the dominant quasinormal modes identified in GW250114~\cite{KAGRA:2025oiz} using a rational filter designed to cancel out their characteristic frequencies \cite{Ma:2022wpv,Ma:2023cwe,Ma:2023vvr}; see Methods for details. Panel (b) of Fig.~\ref{fig:whitened_strain} shows the resulting filtered data, where the quasinormal-mode oscillations are effectively removed. For comparison, we overlay the reconstructed waveform (red) from the numerical-relativity surrogate waveform model NRSur7dq4~\cite{Varma:2019csw}, evaluated at the maximum-a-posteriori parameters of GW250114 inferred from the standard parameter-estimation analysis~\cite{LIGOScientific:2025yae}. Both the original and quasinormal-mode-removed NRSur7dq4 waveforms show close agreement with the observational data.

\begin{figure}[!h]
        \includegraphics[width=\linewidth]{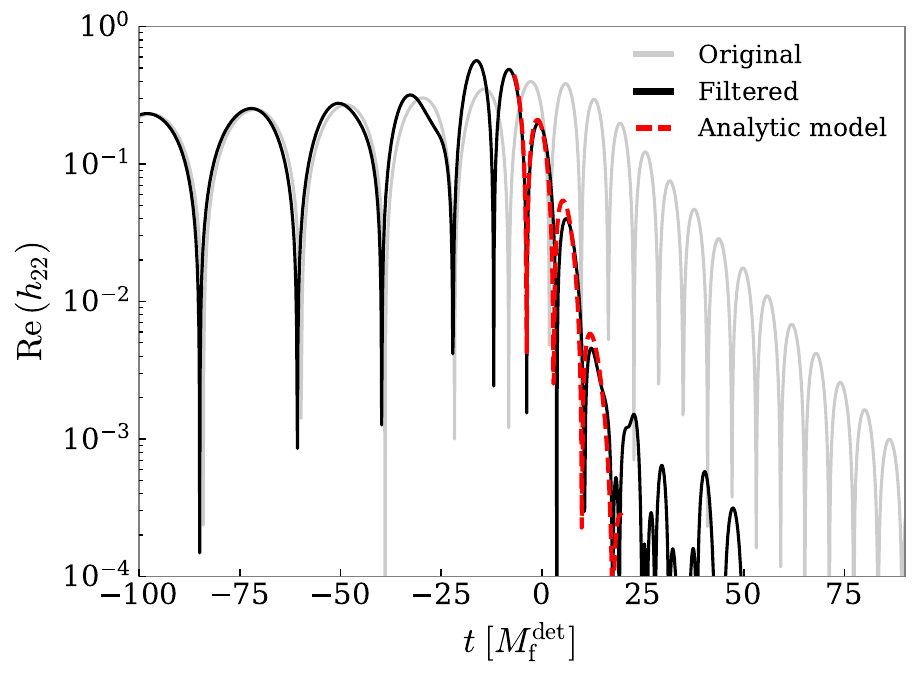}
    \caption{Comparison between waveform components.
    Shown are the real part of the quadrupolar harmonic $(\ell= m =2)$ of the NRSur7dq4 waveform (gray), the waveform after removing the $(\ell= m=2, n=0,1,2)$  quasinormal modes (black), and an analytic model characterizing the direct wave signal (red). The analytic model closely matches the quasinormal-mode-removed waveform as early as $t=-7\Mfdet$.
    }
    \label{fig:theoretical_comparison_7m}
\end{figure}

To illustrate the presence of direct waves around the merger time, we first examine the dominant quadrupolar harmonic $(\ell= m =2)$ in the NRSur7dq4 waveform, shown as the gray curve in Fig.~\ref{fig:theoretical_comparison_7m}. After removing contamination from quasinormal modes with rational filters, the resulting waveform (solid black curve) reveals a few cycles in the interval $t\in [-7,20]\Mfdet$, where time is expressed in units of the remnant black hole mass in the detector frame, with $t=0$ corresponding to the inferred strain peak. We overlay an analytic model for the direct-wave signal (dashed red curve), derived from linear black hole perturbation theory under a near-horizon approximation~\cite{Oshita:2025qmn}:
\begin{equation}
    h_{\rm DW} (t) \sim [\omega_G(t)-\omega_H]  e^{-i \int \omega_G (t) dt }\,,
    \label{eq:analytic}
\end{equation}
where $\omega_G$ is an effective instantaneous complex frequency given by Eq.~(5) in Ref.~\cite{Oshita:2025qmn}, see also the Supplemental Material for more details. As the particle approaches the horizon, $\omega_G$ asymptotes toward $\omega_H$, driving the prefactor $[\omega_G(t) - \omega_H]$ to zero and thereby screening $\omega_H$, which prevents it from being observed at infinity.
The agreement between the quasinormal-mode-removed waveform and the analytic prediction is striking, demonstrating that, once quasinormal modes are removed, the residual oscillations from $t \gtrsim -7\Mfdet$ are clearly identified as the direct-wave signal, which dominates over potential nonlinear or subdominant effects in this regime. 
This provides a compelling theoretical basis for identifying direct waves in observational data. 
Relaxing the near-horizon approximation or including nonlinear effects in the analytical model may extend the agreement to earlier times and refine the description of the onset of direct-wave emission, which we defer to future work.
In what follows, we focus on the time window ($t \gtrsim -7\Mfdet$) to analyze GW250114.

As emphasized in Ref.~\cite{Oshita:2025qmn}, the strong frame dragging in GW250114 causes the direct wave to behave like a damped oscillator with a quasi-stable instantaneous oscillation frequency. This motivates a model-agnostic strategy as a first approximation: we represent the direct wave as a damped sinusoid with constant frequency and damping time, and analyze the quasinormal-mode-removed data within the time-domain Bayesian framework originally developed for quasinormal-mode searches~\cite{Ma:2023cwe,Ma:2023vvr,Lu:2025mwp,Isi:2021iql,Carullo:2019flw}. 
Such a data-driven method minimizes dependence on accurate theoretical templates and provides a robust starting point for identifying the direct-wave signal. 
We note, however, due to the dynamical variations of the instantaneous frequency and damping time of direct waves (see Fig.~5 of Ref.~\cite{Oshita:2025qmn}), our inference should therefore be interpreted as an average frequency and damping rate over a chosen analysis segment, typically $[t_{\rm start}, t_{\rm start}+0.2\,{\rm s}]$ with $t_{\rm start}$ being the analysis starting time. 
To track the slowly evolving features of the direct wave, we repeat the analysis with shifted windows.
The segment length of $0.2\,{\rm s}$ is selected to follow common settings in quasinormal-mode analyses using rational filters~\cite{Lu:2025mwp}. 
We have verified that our results are robust against variations in this choice; see Methods for details.   

\begin{figure}[t!]
    \centering
    \includegraphics[width=\linewidth]{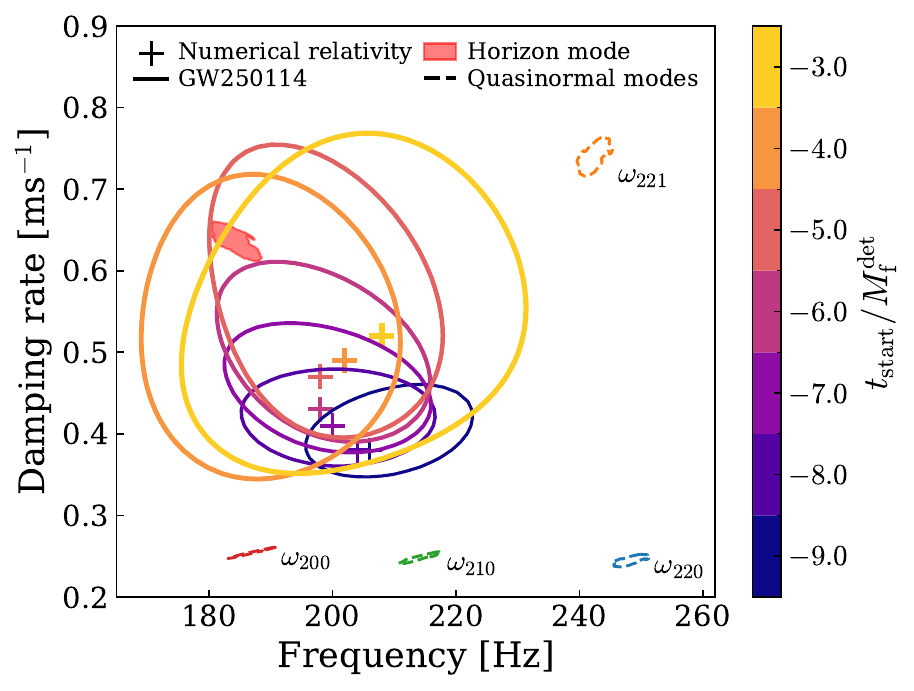}
    \caption{Frequency and damping rate inferred from the data after removing dominant quasinormal modes, revealing the direct-wave signal. Results are shown as a function of analysis starting time (color scale) $\tstart \in [-9,-3]\Mfdet$, covering and extending slightly beyond the theoretically supported regime for direct waves (see Fig.~\ref{fig:theoretical_comparison_7m}). 
    Plus symbols indicate fits to the reconstructed NRSur7dq4 waveform, while colored solid contours represent 90\% credible regions derived from the real event data. Colored dashed contours mark the expected frequencies and damping rates of various quasinormal modes (90\% credible), none of which intersect with the inferred direct-wave parameters. The red shaded region indicates the predicted horizon mode from Eq.~\eqref{eq:omega_H}. While the direct-wave frequencies are not expected to precisely match the horizon mode, they are anticipated to lie nearby. 
    }
    \label{fig:f_tau_fit}
\end{figure}

The 90\% credible posteriors for the average frequency and damping rate inferred from GW250114 are shown in Fig.~\ref{fig:f_tau_fit}, for 
various starting times $\tstart \in [-9,-3]\Mfdet$.
As noted in Fig.~\ref{fig:theoretical_comparison_7m}, the interval $\tstart \gtrsim -7\Mfdet$ corresponds to the regime where the analytic model closely matches the NRSur7dq4 waveform and the direct wave dominates the quasinormal-mode-removed signal. 
We extend the analysis slightly to earlier times ($\tstart \in [-9, -7]\Mfdet$) to explore a possible transition from the inspiral regime to the onset of the direct-wave signal,\footnote{The onset of the direct wave emission is not sharply defined. Relaxing the near-horizon approximation or including nonlinear effects may extend the validity of Eq.~\eqref{eq:analytic} to earlier times; see discussions around Eq.~\eqref{eq:analytic}.} though we caution that fits in this early-time region should not be interpreted as a search for the direct wave.
As a reference, we also fit the same damped sinusoid template to the direct-wave component in the NRSur7dq4 waveform (i.e., the residual cycles shown in the black curve of Fig.~\ref{fig:theoretical_comparison_7m}). The resulting best-fit values are marked by the plus signs in Fig.~\ref{fig:f_tau_fit}. The strong agreement between these fits and the posterior derived from GW250114 data supports the interpretation that our analysis is indeed capturing the genuine direct-wave signal.

For reference, the red-filled contour in Fig.~\ref{fig:f_tau_fit} shows the 90\% credible region for the horizon mode frequency $\omega_H$ [defined in Eq.~\eqref{eq:omega_H}], computed from the remnant mass and spin inferred from the full inspiral-merger-ringdown signal of GW250114. The dashed contours indicate the 90\% credible regions for several relevant quasinormal modes, obtained using the same remnant properties.
Our results show that the extracted average frequency and damping rate evolve and cluster around $\omega_H$ as the analysis window shifts to later times, while remaining clearly distinct from the quasinormal modes.
Because the inferred parameters are dominated by the early portion of each time segment, the frequencies and damping rates shown in Fig.~\ref{fig:f_tau_fit} are expected to reflect the direct-wave evolution and are consistent with the instantaneous behaviour at $t<0$ shown in Supplementary Fig.~\ref{fig:omega_G_evolution}.
The clustering of the oscillation frequency near $2\Omega_H$ reflects the strong frame dragging that forces the orbital motion during the merger to co-rotate with the horizon's intrinsic rotation frequency.
The increasing decay rate traces the continued radial acceleration of the object as it approaches the horizon,
in agreement with the theoretical expectation from Eq.~\eqref{eq:analytic}; see also  
Eq.~(6) in Ref.~\cite{Oshita:2025qmn} and the Supplementary Material.
Finally, the inferred frequency does not fully asymptote to $\omega_H$, due in part to the screening effect of the gravitational potential barrier, and in part to the limited signal-to-noise ratio at late times, which prevents the analysis from extending far enough into the regime where the asymptotic behavior would become resolvable.
Overall, the GW250114 observations closely follow theoretical predictions for direct waves.

To verify that the fitted damped sinusoid is not an artifact of noise, we use a detection statistic $\mathcal{D}$, defined analogously to a logarithmic Bayes factor (see Methods). For an analysis window starting at {$\tstart=-7 \Mfdet$}, the detection statistic comparing the single damped sinusoid model against a noise-only (null) hypothesis yields {$\mathcal{D}=118.1$}. The corresponding false-alarm probability is below 1\% between {$\tstart\in [-7,-3] \Mfdet$}, given a detection threshold of {$\mathcal{D}_{1\%}=1.98$} for a 1\% false-alarm probability. 

Having established the presence and properties of the direct wave using a model-agnostic approach, we next perform a matched-filtering analysis with a more physically motivated analytic template in Eq.~\eqref{eq:analytic}. This waveform enhances sensitivity and sharpens the test of theoretical predictions, further supporting the detection of the direct-wave signal in GW250114. 
The template includes three free parameters: the real and imaginary amplitudes, and a time shift relative to the event data.
We perform the analysis using the same time-domain Bayesian framework~\cite{Isi:2021iql,Carullo:2019flw}; see Methods for details and the inferred posterior distributions. 
Over a 0.2~s segment starting from $-7\Mfdet$, we obtain matched-filter signal-to-noise ratios of $15.8^{+0.1}_{-0.5}$ and $17.1^{+0.1}_{-0.4}$ (90\% credible) in Hanford and Livingston data, respectively, consistent with expectations at Advanced LIGO design sensitivity~\cite{Oshita:2025qmn} (see Table I therein). 
Panel (c) of Fig.~\ref{fig:whitened_strain} shows the residual strain after subtracting the best-fit analytic direct-wave template from the quasinormal-mode-removed data.
The minor dip and surrounding features near $t = 0$ are visual artifacts of the whitening procedure used in the plot, arising from the frequency-dependent rescaling of the data, which exaggerates localized distortions in the time domain; the full direct-wave trough has been effectively removed by the template, leaving only a prominent peak at $t \lesssim -7 \Mfdet$ (see also Fig.~\ref{fig:theoretical_comparison_7m}). 
This early-time peak may also originate from direct-wave emission. A more accurate waveform model is needed to capture its behavior. We leave a detailed investigation of this feature to future work.

\section*{Conclusion}
We analyzed the merger phase of GW250114 as a superposition of free oscillations of quasinormal modes and source-driven direct waves. Using rational filters \cite{Ma:2022wpv,Ma:2023cwe,Ma:2023vvr} as a particularly useful tool, we demonstrated that the recently predicted direct waves \cite{Oshita:2025qmn} can indeed be identified in this event. By modeling the direct waves as a damped sinusoid and a physically motivated waveform template, we found that their measured properties, obtained through time-domain Bayesian inference, align closely with theoretical expectations. 
These results support the interpretation of direct waves as the ``final sound'' directly emitted by the component black holes during their transition from inspiral to merger, just before being silenced by the horizon of the remnant black hole. 

Direct waves encode rich horizon physics, including: (i) strong frame dragging, which forces the waves to oscillate near twice the horizon's rotation frequency $\Omega_H$, largely independent of the binary's inspiral history; (ii) gravitational redshift, which attenuates the signal at a rate governed by the surface gravity $\kappa$; and (iii) additional screening by the surrounding gravitational potential barrier. Together, these features give direct access to two fundamental properties of the black hole horizon; and our analysis of GW250114 provides the first direct observational signatures of both $\Omega_H$ and $\kappa$ in gravitational-wave data.

This study extends the scope of traditional black hole spectroscopy \cite{berti2025} by establishing a complementary channel for probing horizon physics in the dynamical, strong-gravity regime. While quasinormal-mode spectroscopy primarily targets the free oscillations of the remnant, the observation of direct waves opens a novel window into the local near-horizon environment and the frame-dragging dynamics within the ergosphere. 
This new perspective offers exciting avenues for future tests of gravity in the most extreme regime, as well as for exploring the nature of exotic compact objects \cite{Cardoso:2019rvt}. Such tests could be enabled by joint merger-ringdown analyses in which parametrized direct-wave and quasinormal-mode models are fitted simultaneously, allowing consistency tests of remnant and near-horizon properties once theoretical and observational systematics are carefully accounted for.

In this work, we modeled the direct wave using a damped sinusoid and a simplified analytic waveform template, a practical first step, though ultimately insufficient for high-precision analyses. It is therefore timely to develop more accurate waveform templates for direct waves, applicable to precessing binaries and capable of incorporating nonlinear effects. To date, studies of nonlinear ringdown have largely focused on quadratic quasinormal modes \cite{Mitman:2022qdl,Cheung:2022rbm,Ma:2022wpv,Khera:2023oyf,Khera:2024bjs,Ma:2024qcv,Zhu:2024rej,Redondo-Yuste:2023seq,Bucciotti:2024zyp,Bourg:2024jme,Bucciotti:2024jrv,Lagos:2024ekd,Bourg:2025lpd,Ma:2025rnv}; extending these analyses to capture nonlinear contributions to direct waves will be an important direction for future work. 
An equally pressing question is whether direct waves must be accounted for in quasinormal-mode analyses of the post-peak regime~\cite{KAGRA:2025oiz,LIGOScientific:2025obp}. If present at a detectable level, direct waves could bias spectroscopic inferences of black hole parameters or mimic signatures of beyond general relativity physics. 
An additional theoretically interesting direction is to investigate the connection between the direct wave and the dynamical excitation of quasinormal modes, as discussed in Refs.~\cite{Chavda:2024awq,DeAmicis:2025xuh}.
Finally, a systematic search for direct waves across existing gravitational-wave events would enable population-level tests of their universality and parameter dependence, and allow additional features to be identified through cross-event consistency checks~\cite{GWTC-4_results}.

\section*{Acknowledgments}
The authors would like to thank Naritaka Oshita and Huan Yang for fruitful discussions.
This material is based upon work supported by NSF's LIGO Laboratory which is a major facility fully funded by the National Science Foundation.
The authors are grateful for computational resources provided by the LIGO Laboratory and supported by National Science Foundation Grants PHY--0757058 and PHY--0823459.
This research is supported by the Australian Research Council Centre of Excellence for Gravitational Wave Discovery (OzGrav), Project Number CE230100016. 
L.S. is also supported by the Australian Research Council Discovery Early Career Researcher Award, Project Number DE240100206. 
O.J.P. is supported by the Spanish Ministerio de Ciencia, Innovacion y Universidades Ramon y Cajal, RYC2023-044489-I funded by MCIN/AEI/10.13039/501100011033 and the FSE+ and cofinanced by the Universitat de les Illes Balears (UIB). This work was supported by UIB with funds from the Programa de Foment de la Recerca i la Innovació de la UIB 2024-2026 (supported by the yearly plan of the Tourist Stay Tax ITS2023-086); the Spanish Agencia Estatal de Investigación grants RED2024-153978-E, RED2024-153735-E, funded by MICIU/AEI/10.13039/501100011033 and the ERDF/EU; and the Comunitat Autònoma de les Illes Balears through the Conselleria d'Educació i Universitats with funds from the ERDF (SINCO2022/18146). 
Research at Perimeter Institute is supported in part by the Government of Canada through the Department of Innovation, Science and Economic Development and by the Province of Ontario through the Ministry of Colleges and Universities.
Y.C.\ is supported by the Brinson Foundation, the Simons Foundation (Award Number 568762), and by US NSF Grants PHY--2309211 and PHY--2309231.
This manuscript carries LIGO Document No. DCC--P2500608.
\def\bibsection{\section*{References}}
\bibliography{refs}

\section*{Methods}

\subsection{Quasinormal modes}
Quasinormal modes (QNMs) are the oscillatory gravitational waves emitted by a Kerr black hole in response to linear perturbations. They are solutions to the Teukolsky equations, which governs the dynamics of perturbations in Kerr spacetime. Each QNM is labeled by the angular numbers $\ell$ and $m$, and the overtone number $n$. The time evolution of each QNM takes the form of a damped sinusoid with a complex frequency $\omega_{\lmn} = 2 \pi f_{\lmn} - i/\tau_{\lmn}$, where $f_{\lmn}$ is the oscillation frequency and $\tau_{\lmn}$ is the damping time. These complex frequencies depend solely on the mass and dimensionless spin of the perturbed Kerr black hole. 

To probe the presence of direct waves around the merger time, it is essential to first remove the dominant QNMs, which would otherwise contaminate or obscure the features of the direct waves.
The real-valued strain of the ($\ell, m, n$) QNM observed in a gravitational-wave detector is given by
\begin{gather}
    h(t) = A_{\lmn} e^{-(t - t_0)/\tau_{\lmn}} \cos \left[ 2\pi f_{\lmn} (t-t_0) + \phi_{\lmn} \right] \,, \label{eq:QNM_def}
\end{gather}
where $t_0$ is a chosen reference time. Throughout this work, we adopt the convention $f_{\lmn}>0$, with $m>0$ ($m<0$) corresponding to prograde (retrograde) modes~\cite{Berti:2025hly,giesler2024,li2022}.
In Eq.~\eqref{eq:QNM_def} and throughout this paper, we consider only the prograde modes and neglect retrograde modes. The strain in Eq.~\eqref{eq:QNM_def} captures the contribution from two prograde modes, radiating towards the north and south directions relative to the black hole's spin axis.

\subsection{The rational filter}
The rational filter is designed to remove any complex-valued frequency component $\omega'$:
\begin{eqnarray}    
    \mathcal{F}_{\omega'}(\omega) = \frac{\omega - \omega'}{\omega - \omega'^*} \, \frac{\omega + \omega'^*}{\omega + \omega'} \, ,\label{eq:filter_general}  
\end{eqnarray}
where $\omega$ is the real-valued frequency, and $^*$ denotes the complex conjugate. The filtering process is agnostic to the amplitude and phase of the complex-valued frequency component, but is closely related to subtracting their maximum-likelihood estimates under the assumption of white noise in the time domain \cite{Ma:2023cwe,Ma:2023vvr}. 

Previous studies have used the QNM rational filter to remove specific QNMs associated with a black hole of a particular mass and spin~\cite{Ma:2022wpv, Ma:2023cwe, Ma:2023vvr}, i.e., a filter denoted by $\mathcal{F}_{\lmn}(\omega)$ cancels both the mode $\omega_{\lmn}$ and its mirroring counterpart $-\omega^*_{\lmn}$. 
If multiple QNMs are present, their combined contribution can be removed by applying the total filter:
\begin{equation}
\mathcal{F}_\text{tot}(\omega) = \prod_{\lmn} \mathcal{F}_{\lmn}(\omega) \, . \label{eq:total_filter}
\end{equation}
We apply the QNM rational filter, constructed using the remnant black hole mass and spin inferred from the full signal, to pre-process the data by removing the dominant QNM contributions prior to searching for direct waves. 

Since the filter is applied in the frequency domain, we first transform the gravitational-wave time-series data, $d_n$, where $n$ indexes discrete time samples, into the frequency domain (denoted $\tilde{d}$) using a discrete Fourier transform.
After applying the frequency-domain filter to $\tilde{d}$, the resulting data are transformed back into the time domain via an inverse discrete Fourier transform, yielding the filtered time series $d^F_n$, where the superscript ``$F$'' denotes the filtered data. 

The effect of applying a QNM filter $\mathcal{F}_{\lmn}$ to low-frequency components is equivalent to introducing a time shift \cite{Ma:2022wpv}:
\begin{equation}    
    t_\lmn = \frac{4}{\tau_\lmn |\omega_\lmn|^2},
    \label{eq:time_shift}
\end{equation}
along with an associated phase shift. We do not write the phase shift explicitly, as it does not impact our analysis. We have verified that, for both inspiral and direct-wave signals, Eq.~\eqref{eq:time_shift} provides an accurate correction for the time shift induced by removing the QNMs. This is illustrated in Fig.~\ref{fig:whitened_strain}, which shows good agreement between the original and filtered data in the inspiral phase after applying the correction from Eq.~\eqref{eq:time_shift}.

\subsection{Quasinormal mode removal}

The ringdown spectrum of GW250114 has been thoroughly studied in Ref.~\cite{LIGOScientific:2025obp}, which reports strong evidence for the $(\ell=m=2,n=0,1)$ QNMs, with weak early-time preference for the presence of the $(\ell=m=2,n=2)$ mode at $t \lesssim 5\Mfdet$. Meanwhile, late-time tails are expected to be negligible in our scenario \cite{Ma:2024hzq,DeAmicis:2024eoy,Islam:2024vro,DeAmicis:2024not,Ma:2024bed,Cardoso:2024jme}.

Throughout this work, we first apply the QNM rational filter to remove these three QNMs prior to searching for the direct-wave signal. 
In Ref.~\cite{LIGOScientific:2025obp}, the fundamental $\ell = m = 4$ mode is constrained to within tens of percent by fitting a parameterized waveform that models the full inspiral–merger–ringdown sequence. However, the signal-to-noise ratio of the $(\ell=m=4,n=0)$ mode is insufficient for detection through any standalone ringdown analysis. As such, we do not explicitly filter out the $(\ell=m=4,n=0)$ QNM in our direct wave analysis.
The filters are constructed using the maximum-a-posteriori values of the (detector-frame) final black hole mass and spin. The time shift induced by applying the filters on the remaining signal is corrected using Eq.~\eqref{eq:time_shift}.

\subsection{Model-agnostic search for direct waves}

After removing the dominant QNMs from the original data and correcting for the time shift, we search for direct waves as a damped sinusoid with an unknown complex-valued frequency $\omega'$ and define the likelihood function as \cite{Ma:2023cwe,Ma:2023vvr}
\begin{gather}
    \text{ln } \mathcal{L}(d |\omega', \tstart) = -\frac{1}{2} \sum_{i,j>0} d_i^F C_{ij}^{-1} d_j^F \, , \label{eq:likelihood}
\end{gather}
with $\tstart$ denoting the starting time of the analysis, and $d$ is taken over an interval of $[\tstart, \tstart+0.2\,{\rm s}]$. The matrix $C_{ij}$ is the autocovariance of the detector noise, estimated using the Welch method~\cite{welch1967} over 64~s of data starting 1~s after the reference time at the inferred strain peak, during which no gravitational wave signals were identified, but the noise characteristics are expected to closely match those during the event.

The QNM-removed data are analyzed using the free-frequency filter defined in Eq.~\eqref{eq:filter_general}. Given the low computing cost of the rational filter, we efficiently evaluate the likelihood in Eq.~\eqref{eq:likelihood} over a grid of real-valued frequencies and damping times, assuming uniform priors on the frequency (160--240~Hz) and damping rate (0.05--0.9~ms$^{-1}$). The grid is constructed with a resolution of 2~Hz in frequency and 0.01~ms$^{-1}$ in damping rate. 

The analysis is performed using multiple choices of starting times. For each selected start time, a data segment of duration $L_\mathrm{seg} = 0.2$~s is analyzed. The data are sampled at a rate of $f_\mathrm{samp}=8192$ Hz. The LIGO Hanford and Livingston detector data are combined incoherently after correcting for the signal travel time between the detectors, using the maximum-a-posteriori sky localization and geocentric signal start time inferred from the inspiral-merger-ringdown parameter estimation~\cite{KAGRA:2025oiz,LIGOScientific:2025obp, GW250114_082203_2025}. 

\begin{figure}
        \includegraphics[width=\linewidth]{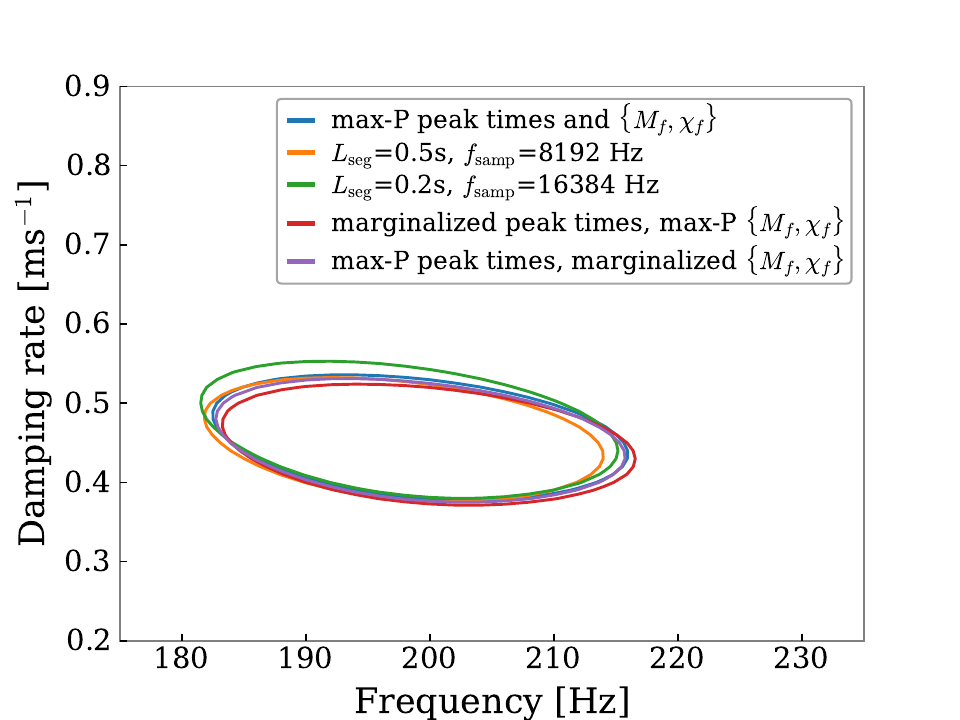}
    \caption{Inferred frequency and damping rate (90\% credible region) of the direct-wave signal at $\tstart = -7\Mfdet$ under different analysis settings for $L_\mathrm{seg}$, $f_\mathrm{samp}$, strain peak times, and remnant black hole properties $\{M_f, \chi_f\}$ (with ``max-P'' denoting maximum-a-posteriori values). 
    The red (purple) contour marginalizes over the strain peak time in each detector (remnant black hole parameters). The timing uncertainties reflect variations in geocentric peak time and sky location, while the remnant-parameter marginalization accounts for possible variations in the QNM $(\ell=m=2,n=0,1,2)$ frequencies.
    The main results presented in the paper correspond to the settings in blue. The analysis is robust against variations in these settings.}
    \label{fig:checks}
\end{figure}

The robustness of these analysis settings is verified, as shown in Fig.~\ref{fig:checks}. The inferred frequencies and damping rates of the direct-wave signal (at start time $\tstart = -7 \Mfdet$) are insensitive to variations in $L_\mathrm{seg}$, $f_\mathrm{samp}$, as well as to the choice of strain peak time in each detector (equivalently parameterized by sky location and geocentric signal start time), and to the remnant black hole mass and spin used in the quasinormal-mode subtraction. 
To demonstrate robustness with respect to the signal timing and remnant properties, we draw 200 samples in each case from the inspiral-merger-ringdown posterior distribution, marginalize over these draws, and plot the resulting contours. In both cases, the marginalized contours remain consistent with the main analysis.\footnote{Ref.~\cite{siegel2025} provides a more rigorous treatment of the necessary conditions for verifying that analysis settings do not bias the inference results. Here, we qualitatively verify that our analysis is unaffected by the analysis settings and leave a more quantitative analysis to future work.}

To quantify the preference for a direct-wave signal model over one containing pure noise after filtering out the QNMs from the data, we adopt a hybrid Bayesian-like approach~\cite{Lu:2025mwp}.
The detection statistic $\mathcal{D}$, analogous to a logarithmic Bayes factor,\footnote{It formally differs from a Bayes factor computed using methods that also compute the amplitude and phase of the complex-valued signal (see Appendix A in Ref~\cite{Lu:2025mwp}).} is defined as a comparison between two model hypotheses: $\mathcal{H}$, which includes a direct-wave content, and $\mathcal{H'}$, which does not. Given data $d$, the statistic is:
\begin{equation}
\mathcal{D}(\mathcal{H}:\mathcal{H}') = \log_{10}\frac{\mathcal{Z}(d | \mathcal{H})}{\mathcal{Z}(d | \mathcal{H'})},
\end{equation}
where $\mathcal{Z}(d|\mathcal{H})$ and $\mathcal{Z}(d|\mathcal{H'})$ denote the evidences under hypotheses $\mathcal{H}$ and $\mathcal{H'}$, respectively.
To assess the statistical significance of a detection, we adopt a frequentist approach to estimate false alarm probabilities due to the background noise. 
We determine the detection threshold by performing injection studies in which a QNM-only signal $(\ell=m=2,n=0,1,2)$ is added to the detector noise surrounding the event. 
These QNMs are injected with amplitudes drawn uniformly from $4\times10^{-22}$ to $2\times10^{-21}$ and phases from 0 to $2\pi$ at the reference time, while their complex frequencies are fixed to the maximum-a-posteriori remnant parameters inferred from the inspiral-merger-ringdown analysis.
For each simulated realization, we compute the detection statistic comparing a model that includes an additional damped sinusoid (representing a direct-wave component) against a baseline model in which only the three injected QNMs are removed, with their parameters fixed to the injected values. 
The prior ranges and parameter resolution for the additional damped-sinusoid search are identical to those used in the direct-wave search of the event data.
This setup therefore mirrors our search for an additional damped sinusoid on top of the three QNMs with parameters determined from the inspiral-merger-ringdown analysis.
Repeating this procedure for 300 independent noise realizations (drawn at 64-s intervals within the 3 hours before and after the event, excluding the 10 s surrounding the event itself) yields the distribution of the detection statistic under QNM-only null hypothesis. We define the threshold $\mathcal{D}_{1\%}$ as the value corresponding to a 1\% false-alarm probability.
An observed $\mathcal{D} -\mathcal{D}_{1\%}>0$ therefore indicates that the preference for a direct-wave signal is unlikely to arise from noise fluctuations alone. 

\subsection{Template-based search for direct waves}
\label{methods:template_search}
\begin{figure}[!h]
        \includegraphics[width=\linewidth]{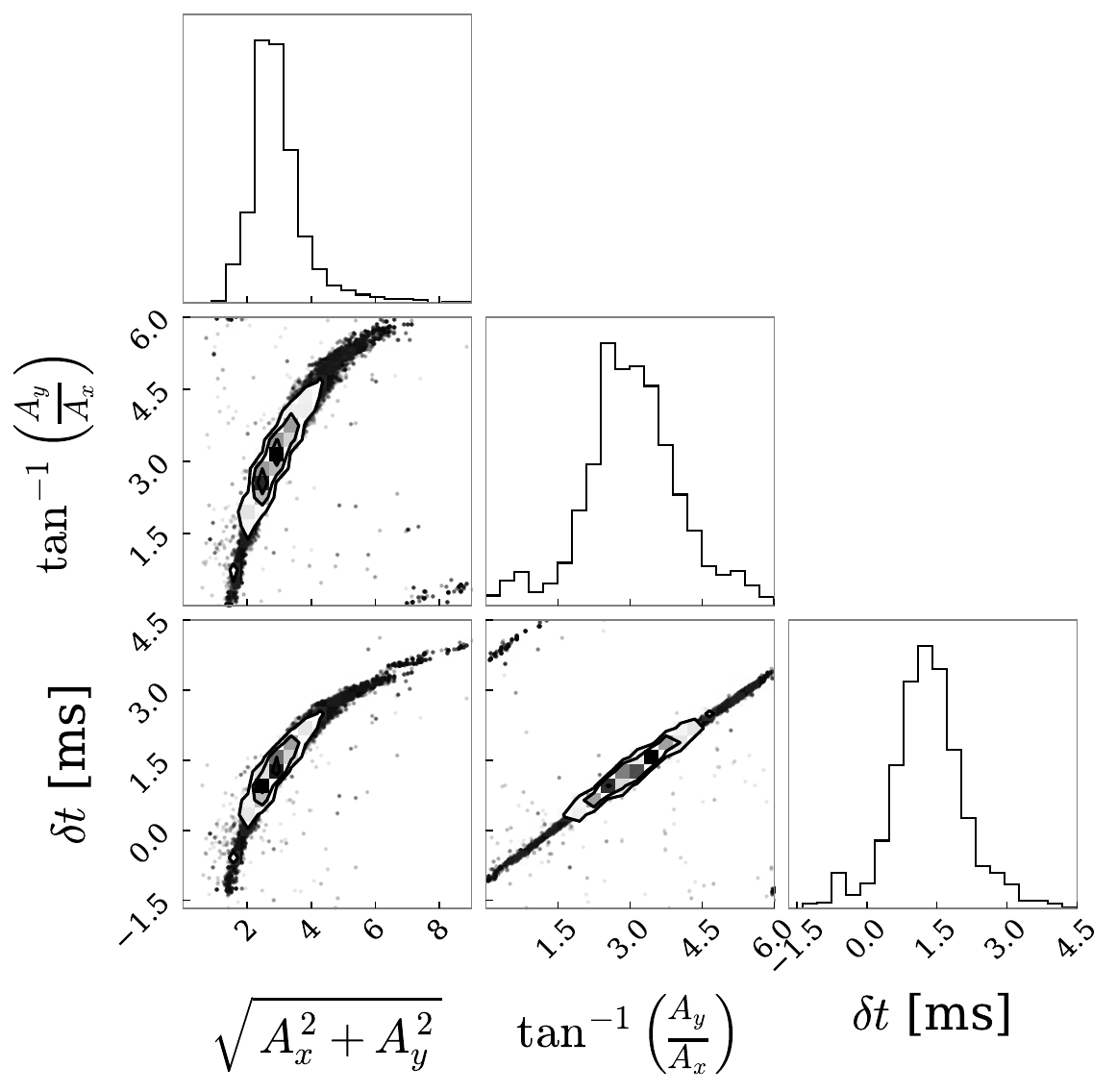}
    \caption{Posterior distribution of the free parameters $\left\{\sqrt{A_x^2+A_y^2}, \tan^{-1}\left(\frac{A_y}{A_x}\right), \delta t\right\}$ obtained from the template-based search.}
    \label{fig:corner_plot}
\end{figure}

We perform matched-filtering analyses based on the analytic direct-wave template defined in Eq.~\eqref{eq:analytic}.
To construct the signal template $h_{\rm temp}(t)$ in the detector frame, we first benchmark the analytic waveform $ h_{\rm DW} (t)$ in Eq.~\eqref{eq:analytic} against the NRSur7dq4 waveform (see Fig.~\ref{fig:theoretical_comparison_7m}), evaluated at the maximum-a-posteriori remnant parameters inferred for GW250114 (remnant mass, spin, and luminosity distance), by applying a time shift and fitting a single complex-valued amplitude. This procedure yields the benchmarked waveform $h_{\rm bm}(t)$.
We then introduce three free parameters: a complex amplitude, $A_x+iA_y$, to rescale $ h_{\rm bm} (t)$, and a time shift $\delta t$ relative to the event data, such that the quadrupolar strain waveform takes the form
$(A_x+iA_y)\times h_{\rm bm} (t-\delta t)$, 
which is subsequently projected to the detector-frame to obtain $h_{\rm temp}(t)$, using the maximum-a-posteriori extrinsic parameters (inclination angle, sky location, and polarization angle). 
Finally, we fit $h_{\rm temp}(t)$ separately to the LIGO Hanford and Livingston data, after removing the QNMs.
The fit is performed within the time-domain Bayesian framework~\cite{Isi:2021iql,Carullo:2019flw}, using a 0.2~s window starting at $t = -7 \Mfdet$, where the direct-wave signal is expected to emerge. Uniform priors are adopted for $A_x$ and $A_y$ over the range $[-20, 20]$, and for $\delta t$ over the interval $[-20,20]\,\Mfdet$.
The posterior distribution is shown in Fig.~\ref{fig:corner_plot}.
The matched-filter signal-to-noise ratio is defined as
\begin{equation}
    \rho_{\rm mf} = \frac{\langle d_n^F | h_{\rm temp}\rangle}{\sqrt{\langle h_{\rm temp} | h_{\rm temp}\rangle}},
\end{equation}
where $\langle \cdot|\cdot \rangle$ denotes the usual noise-weighted inner product in the time domain. We compute the recovered signal-to-noise ratio from the Hanford and Livingston detectors independently.

\begin{figure}[!h]
        \includegraphics[width=\linewidth]{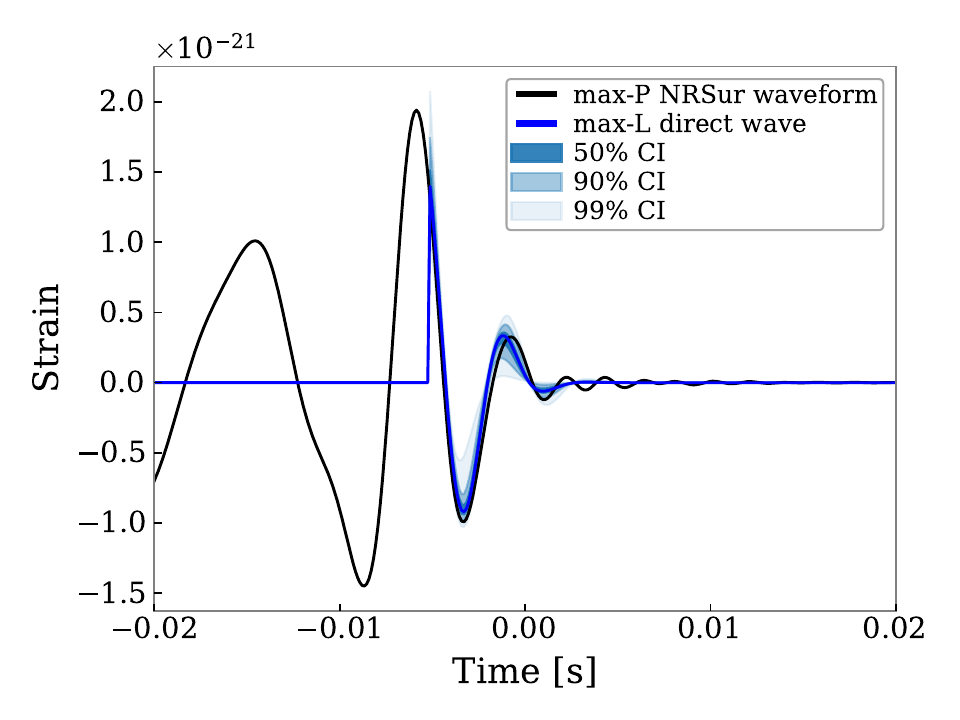}
    \caption{Waveform reconstruction. Black: maximum-a-posteriori (``max-P'') NRSur7dq4 waveform from the inspiral-merger-ringdown analysis, with the $(\ell=m=2,n=0,1,2)$ QNMs removed. Blue: maximum-likelihood (``max-L'') direct-wave reconstruction. Shaded bands: 50\%, 90\%, and 99\% credible intervals.}
    \label{fig:reconstruct_dw}
\end{figure}

We show the waveform reconstruction in Fig.~\ref{fig:reconstruct_dw}. The NRSur7dq4 waveform (``max-P'') is constructed from the maximum-a-posteriori inspiral-merger-ringdown parameters and preprocessed identically to the data, with the $(\ell=m=2,n=0,1,2)$ QNMs removed. It is in good agreement with the direct-wave reconstruction (``max-L'') obtained from a maximum-likelihood fit of the analytical template. The shaded regions indicate the 50\% and 90\% credible intervals derived from the posterior distribution of the direct-wave parameters. Residual oscillations at late times in the NRSur7dq4 waveform arise from a subdominant $(\ell=m=4,n=0)$ QNM with negligible signal-to-noise ratio and do not affect the fit.

\section*{Supplemental Material}
\setcounter{subsection}{0}

\subsection{Near-horizon wave emission and horizon modes}
\label{supp:horizon_mode}

{
\begin{table*}
\centering
 {   \begin{tabular}{c|c|c|c|c|c|c}
 &    Coordinates & Horizon & Particle & Observer & Outgoing rays & Features\\
    \hline
 (a) &    \begin{tabular}{c}
    Boyer–Lindquist \\ $(t,r)$
    \end{tabular} &
    \begin{tabular}{c}
 future/past horizons \\
    both at $r=r_+$
    \end{tabular}
    & 
    \begin{tabular}{c}
     asymptote to $r_+$ \\ 
    in the future 
    \end{tabular}
    & \begin{tabular}{c}
    constant \\ position
    \end{tabular}& \begin{tabular}{c}
    asymptote to $r_+$ in the \\
    past, $45^\circ$ in the future
    \end{tabular}
    & \begin{tabular}{c}
    particle and rays \\
    freeze on the horizon
    \end{tabular}
    \\
    \hline
        (b) &
        \begin{tabular}{c} 
        Tortoise \\ $(t,r_*)$ 
        \end{tabular}
        &
    outside range & 
    \begin{tabular}{c}
    nearly $45^\circ$ \end{tabular}
    & \begin{tabular}{c}
    constant \\ position
    \end{tabular}& $45^\circ$ &
\begin{tabular}{c}   near-horizon motions \\
simplified
\end{tabular}
\\ \hline
(c) & \begin{tabular}{c} Ingoing EF \\
$(t_{\rm in},r)$ \end{tabular} & 
    \begin{tabular}{c}
    future/past horizons \\
    both at $r=r_+$
    \end{tabular}
    & 
    \begin{tabular}{c}
    penetrates \\ horizon
    \end{tabular}
    & \begin{tabular}{c}
    constant \\ position
    \end{tabular}& \begin{tabular}{c}
    asymptote to $r_+$ in the \\
     past, $45^\circ$ in the future
    \end{tabular} &
\begin{tabular}{c}
    regular near the \\ 
    future horizon
    \end{tabular}
    \\
    \hline
(d) &
\begin{tabular}{c}
Kruskal  \\
$(U,V)$
\end{tabular}
& 
\begin{tabular}{c}
All resolved \\ at $UV = 0$ 
\end{tabular}
& 
    \begin{tabular}{c}
    penetrates \\ horizon
    \end{tabular}
&     \begin{tabular}{c}
    hyperbolic \\ worldline
    \end{tabular}
    & $45^\circ$ & 
regular at all horizons
    \end{tabular}}
    \caption{Comparison between coordinate systems. }\label{table1}
\end{table*}}

Direct waves and horizon modes are outgoing waves emitted near a black hole's horizon. The appearance of this emission, however, depends on the observer's frame. Below, we take a pedagogical approach to illustrate how different observers perceive this process.

\begin{figure}
    \includegraphics[width=0.5\textwidth]{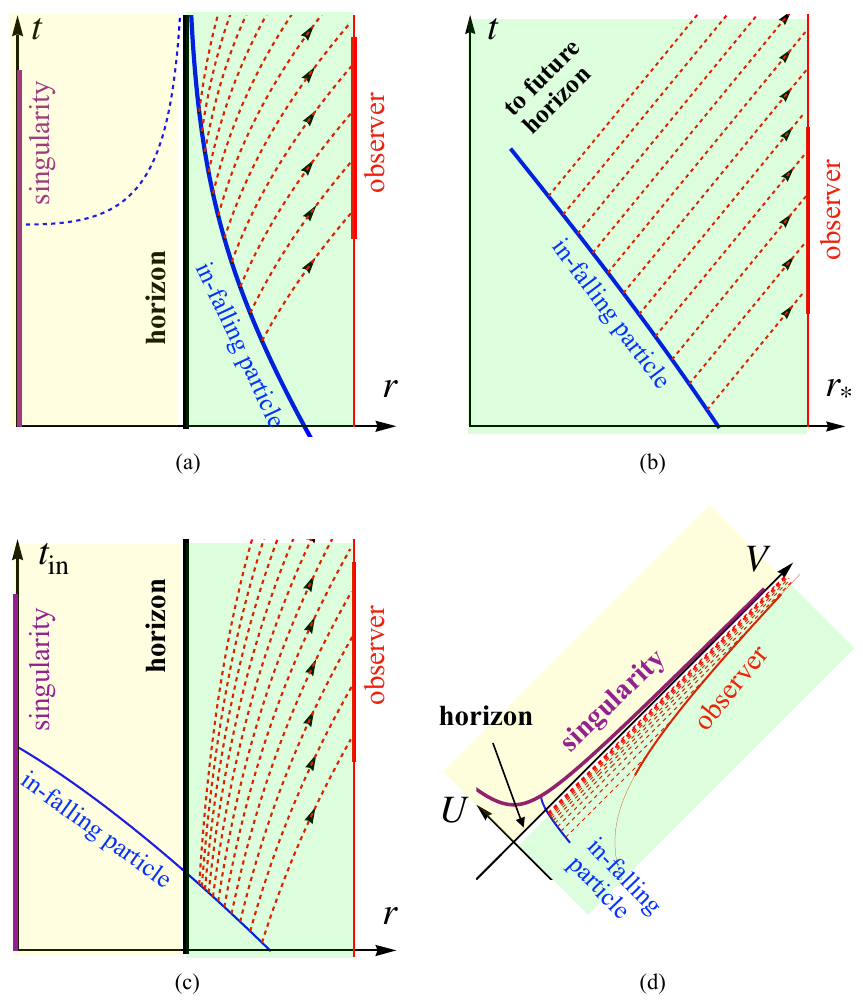}
    \caption{Four coordinate systems depicting the same particle falling into a black hole and crossing the horizon in finite proper time. The particle emits rays at exponentially decreasing proper-time intervals before horizon crossing. (a) In Boyer–Lindquist coordinates, the particle asymptotes to the horizon without reaching it, and the emitted rays appear to be spread out over all times. (b) In the tortoise coordinate system, which is well adapted for illustrating wave propagation, the horizon is at $r_*\rightarrow -\infty$; the ingoing trajectory of the particle approaches $45^\circ$ with $dr_*/dt =-1$. The emitted outgoing rays satisfy $dr_*/dt =1$. (c) In ingoing Eddington–Finkelstein coordinates, the particle smoothly crosses the horizon, and the shrinking proper-time intervals between ray emissions before horizon-crossing are evident. (d) In Kruskal coordinates that maximally extend a Kerr spacetime, the particle follows an approximately constant-$V$ trajectory, while the outgoing rays follow constant-$U$ lines.
    \label{fig:coords}  }
\end{figure}

\begin{figure}
    \includegraphics[width=0.475\textwidth]{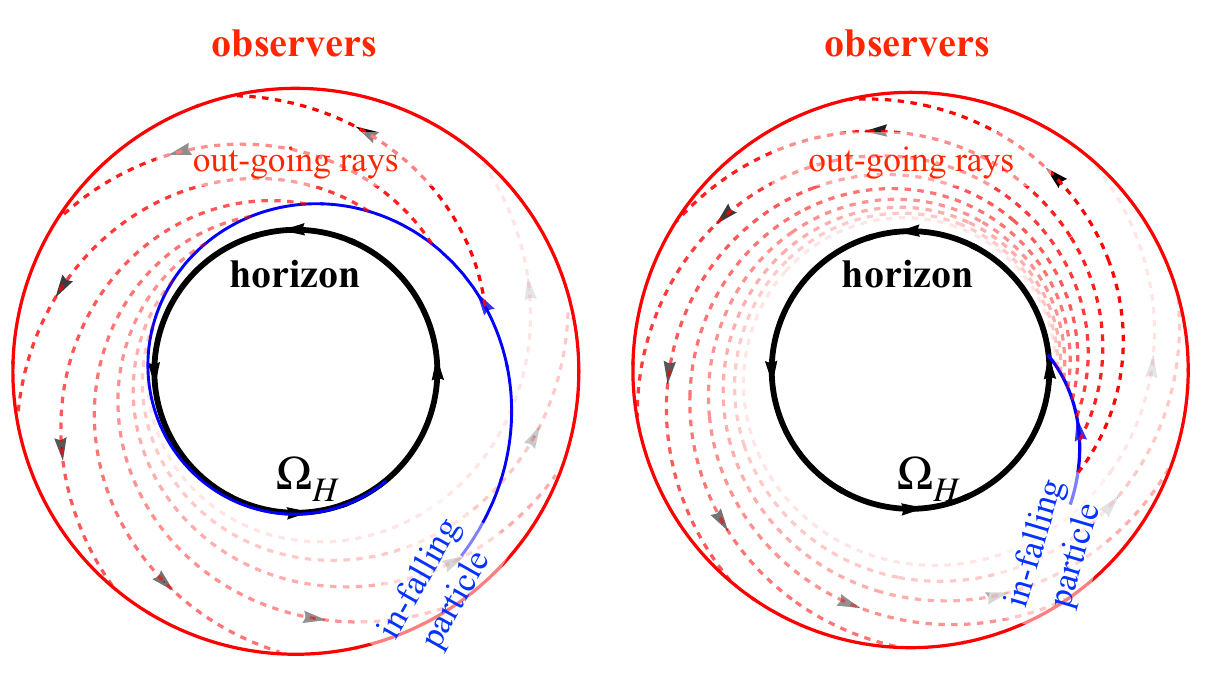}
    \caption{Frame-dragging in the near-horizon region of a Kerr black hole, whose horizon rotates at angular frequency $\Omega_H$ relative to distant observers. The left and right panels show top views of the coordinate systems in Fig.~\ref{fig:coords} (a) and (c), respectively, plotted using the azimuthal angle measured by distant observers.
    \label{fig:kerr_rays} }
\end{figure}

To formulate this more precisely, we introduce four coordinate systems (see Table~\ref{table1}), each corresponding to a family of observers whose spatial locations are arranged in a specific manner. Every observer carries a clock, and these clocks are synchronized and tick according to the time coordinate of their respective system. 
We describe the same process in alternative coordinate systems, each offering a distinct observational perspective.
The Boyer--Lindquist coordinates $(t,r,\theta,\phi)$ provide perhaps the most familiar description, as they align closely with our intuitive notions of spacetime decomposition and serve as the natural coordinate system for distant observers.
Fig.~\ref{fig:coords} (a) shows a particle falling down a black hole viewed by Boyer–Lindquist observers. The particle asymptotically approaches the horizon at $r_+$
and appears to freeze there, never actually crossing it. Rays emitted closer to the horizon take progressively longer to reach a finite distance away from the black hole; as a result, they are increasingly stretched (redshifted) at late times. Meanwhile, the left panel of Fig.~\ref{fig:kerr_rays} shows the corresponding azimuthal motion: rays emitted near the horizon swirl more angles before propagating outward.

The radial motion is described more naturally using the tortoise radial coordinate $r_*$:
\begin{align}
    r_*=\int \frac{r^2+\chi^2}{\Delta} dr,
\end{align}
with $\Delta=r^2-2r+\chi^2$. This coordinate maps the near-horizon region to an unbounded domain ($r=r_+$ is mapped to $r_*=-\infty$), which naturally encodes the near-horizon gravitational redshift effect and thus provides a more suitable way to describe the propagation of outgoing rays. Indeed, as illustrated in Fig.~\ref{fig:coords} (b), rays that are equally spaced in $r_*$ remain equally spaced as they propagate outward. 
These rays, on the other hand, are emitted at exponentially decreasing intervals with respect to the emitter's proper time. Specifically, the emitter's motion near the horizon satisfies, see i.e., Eq. (2.4) in \cite{Mino:2008at}
\begin{align}
    \frac{dr}{d\tau}\propto {\rm const},
\end{align}
which implies that the emitter's proper time $\tau$ is proportional to $r\sim r_++e^{2\kappa r_*} $. Consequently, a sequence of rays that are equally spaced in $r_*$ (and thus received at equal time intervals by a distant observer) are emitted at exponentially shrinking proper-time intervals $\Delta\tau$ by the infalling source.

The ingoing Eddington-Frankestein coordinate system $(t_{\rm in},r,\theta,\phi_{\rm in})$ is defined as:
\begin{subequations}
\begin{align}
&t_{\rm in}=t+r_*-r, \\
&\phi_{\rm in}=\phi+\int \frac{\chi}{\Delta}dr,
\end{align}
\end{subequations}
where the coordinate time $t_{\rm in}$ corresponds to the clocks carried by observers whose ticking rate depends on the radial position. In this coordinate system, ingoing principal null geodesics trace straight coordinate lines,
\begin{align}
t_{\rm in}+r={\rm const},
\end{align}
with no azimuthal motion, i.e., $\phi_{\rm in}={\rm const}$. The ingoing Eddington–Finkelstein coordinates are therefore well suited for describing trajectories that fall into the black hole, such as the plunging motion relevant to binary coalescences. An example of such a trajectory is shown by the blue curve in Fig.~\ref{fig:coords} (c), which reaches the horizon at a finite coordinate time $t_{\rm in}$ and smoothly crosses it. Outgoing rays emitted along the way (red dashed curves) satisfy
\begin{align}
    \frac{dt_{\rm in}}{dr}= 2\frac{r^2+a^2}{\Delta}-1,
\end{align}
indicating that the closer the emission occurs to the horizon, the longer it takes for the rays to escape to distant observers. Similarly, the angular evolution of outgoing rays is given by
\begin{align}
    \frac{d\phi_{\rm in}}{dt_{\rm in}}=\frac{2\chi}{2(r^2+\chi^2)-\Delta}.
\end{align}
The right panel of Fig.~\ref{fig:kerr_rays} illustrates the azimuthal motion: rays that take longer to propagate outward undergo more angular winding around the black hole.

Finally, the Kruskal coordinate system $(U,V,\theta,\tilde{\phi})$ is defined as \cite{o2014geometry}
\begin{subequations}
\begin{align}
&U = -e^{-\kappa (t-r_*)}, \label{supp:eq:Kruskal_U} \\
&V=e^{\kappa (t+r_*)}, \\
&\tilde{\phi}=\phi-\Omega_H t. \label{supp:eq:Kruskal_phi}
\end{align}
\end{subequations}
As illustrated in Fig.~\ref{fig:coords} (d), Kruskal observers correspond to the ingoing principal null geodesics ($V={\rm const}$) that fall into the black hole horizon and the outgoing principal null geodesics ($U={\rm const}$) that emerge from the white hole horizon. This Kruskal system remains regular across the horizons and provides a well-behaved description of the maximal extension of the Kerr spacetime around the bifurcation sphere $(U=V=0)$. The blue curve in Fig.~\ref{fig:coords} (d) represents a plunging trajectory. As the particle accelerates to nearly the speed of light near the horizon, its path becomes approximately aligned with a constant$-V$ line and crosses the horizon $(U=0)$ smoothly. The emitted outgoing rays form straight coordinate lines $U={\rm const}$ (red dashed lines). A distant observer, located at a fixed radius $r$, follows a worldline satisfying $UV={\rm const}$ (red solid curve). The time intervals between the received signals are stretched (redshifted) at late times.

The outgoing waves emitted by the plunging particle can be represented by an analytic function $\psi$ of the retarded time $U$ and $\tilde{\phi}$, whose Taylor expansion reads
\begin{align}
    \psi(U,\tilde{\phi})  \sim \sum_{n,m} c_{n,m}U^n e^{im\tilde{\phi}},
\end{align}
where $n$ and $m$ are integers, and $c_{n,m}$ are coefficients.
The corresponding signals received by the distant observer follow the same functional dependence and can be interpreted as a superposition of an infinite set of horizon modes 
\begin{align}
    U^n e^{im\tilde{\phi}}\sim e^{-i \omega^{(mn)}_H t}e^{im\phi},
\end{align}
with
\begin{equation}
    \omega^{(mn)}_H = m\Omega_H -i n\kappa. \label{supp:eq:HM}
\end{equation}
Here we have used Eqs.~\eqref{supp:eq:Kruskal_U} and \eqref{supp:eq:Kruskal_phi}.

Another feature of horizon modes emerges from the near-horizon geometry. Using the Rindler coordinates
\begin{equation}
    dt^2 = -\kappa^2 \rho^2 dt^2 + d\rho^2 + \gamma_{AB}dX^A dX^B\,,
\label{metric:NH}
\end{equation}
with $\rho \approx \sqrt{2(r-r_+)/\kappa}\propto e^{\kappa r_*}$,
the metric can be Euclideanized via a Wick rotation $t = i\tau$. Identifying $\tau \rightarrow \tau +2\pi \kappa$ makes a locally smooth space free from conical singularity~\cite{gibbons1977action}.  Horizon modes are single-valued smooth functions in this Euclidean space, while other modes are multi-valued.

\subsection{Screening of the horizon modes}
In Sec.~\ref{supp:horizon_mode}, we show that the horizon modes, $\omega^{(mn)}_H$ defined in Eq.~\eqref{supp:eq:HM}, represent outgoing waves emitted near the horizon. However, here we argue that these modes are entirely screened by the black hole potential barrier and do not contribute to the direct waves observed at infinity. 

The most direct argument follows Refs.~\cite{Oshita:2025qmn,ZMC18}, which explicitly show that the black hole's greybody factor, the transmissivity of outgoing waves emitted from the past horizon toward future null infinity, vanishes exactly at $\omega^{(mn)}_H$. This property has also been discussed in the context of the Kerr/CFT correspondence~\cite{castro2013black,bredberg2010black} and in classical literature on black hole perturbation theory. For example, the energy absorption coefficient $\Gamma$ in Eq.~(4.7) of Ref.~\cite{Mano:1996vt} vanishes at poles of $A_{\rm in}^{s\nu}A_{\rm in}^{-s\nu}$, which, according to their Eq.~(4.2), includes $m\Omega_H -i (1+n\pm s) \kappa$ with $s$ the spin weight of the field and $n=0,1,2,\ldots$. 

Another instructive way to understand the screening effect is by considering the Hawking flux $\mathcal{N}_{\ell m}(\omega)$ at infinity for the $(\ell, m)$ mode:
\begin{equation}
\label{HR_flux}
    \mathcal{N}_{\ell m}(\omega) = \frac{\Gamma_{\ell m}(\omega)}{\exp[\beta(\omega-m \Omega_H)]-1}\,,
\end{equation}
where $\Gamma_{\ell m}(\omega)$ is the corresponding energy greybody factor, and $\beta = 2\pi / \kappa$ is the inverse temperature of the black hole. The Bose–Einstein factor in the denominator has poles precisely at the horizon mode frequencies $\omega_H^{(mn)}$. In this sense, the horizon modes can be interpreted as the Matsubara frequencies of the black hole~\cite{Matsubara1955}, corresponding to those with negative imaginary parts. The vanishing of $\Gamma_{\ell m}(\omega)$ at $\omega_H^{(mn)}$ makes the flux $\mathcal{N}_{\ell m}(\omega)$ remain finite by canceling the divergences at these poles. 

\subsection{Direct waves}
Although the horizon modes $\omega^{(mn)}_H$ are completely screened by the black hole potential barrier~\cite{Oshita:2025qmn,ZMC18}, direct waves emitted not too close to the horizon can still reach distant observers, as noted in Ref.~\cite{Oshita:2025qmn}. In the following, we briefly summarize the derivation of the analytic waveform for the direct waves used in the main text, and refer the reader to \cite{Oshita:2025qmn} for further details.

The gravitational-wave strain can be obtained by two time integrations of the Weyl scalar $\Psi_4$. Using the Teukolsky equation \cite{Teukolsky:1973ha,PhysRevLett.29.1114}, the asymptotic form of $\Psi_4$ observed at infinity can be written as
\begin{align}
    [r\Psi_4 (u)]_{r\to\infty,\ell m}= \int d\omega e^{-i\omega u} Z_{\ell m\omega}, \label{eq:SI:Psi4}
\end{align}
where $u$ is the retarded time, and $\Psi_4$ has been decomposed into spin-weighted spheroidal harmonics labeled by $(\ell, m)$. The quantity $Z_{\ell m\omega}$ denotes the Fourier mode at frequency $\omega$, given by
\begin{align}
    Z_{\ell m\omega}=\frac{1}{2i\omega B^{\rm in}_{\ell m\omega}}\int dr \frac{R^{\rm in}_{\ell m\omega}S_{\ell m \omega}}{\Delta^2}, \label{eq:SI:Z}
\end{align}
where $\Delta=(r-r_+)(r-r_-)$ and $r_\pm$ denote the radii of the event and Cauchy horizons, respectively. $S_{\ell m \omega}$ is the source term contributed by the perturber. $R_{\ell m\omega}^{\rm in}$ is the ingoing homogeneous solution to the Teukolsky equation, satisfying the boundary conditions
\begin{align}
    R_{\ell m\omega}^{\rm in}(r) = 
\begin{cases} 
 \Delta^2 e^{-ikr_*} & r \to r_+ \\ 
B_{\ell m\omega}^{\text{out}} r^3 e^{i\omega r_*} + r^{-1} B_{\ell m\omega}^{\text{in}} e^{-i\omega r_*} & r \to \infty
\end{cases},
\end{align}
with $k=\omega-m\Omega_H$.

Here we focus on the contribution from waves sourced in the near-horizon region. Specifically, we restrict the integration in Eq.~\eqref{eq:SI:Z} to $r \sim r_+$, under which the expression simplifies to \cite{Mino:2008at,Zimmerman:2011dx,Oshita:2025qmn}
\begin{align}
    Z_{\ell m\omega}=\hat{D}_{\ell m\omega}\tilde{Z}_{\ell m\omega} \int dt e^{i\omega t-im\phi(t)} e^{-i (k+2i\kappa)r_*(t)}, \label{eq:SI:near_horizon_limit}
\end{align}
where $\phi(t)$ and $r_*(t)$ denote the orbital phase and tortoise radial coordinate of the perturber, respectively. The explicit forms of $\hat{D}_{\ell m\omega}$ and $\tilde{Z}_{\ell m\omega}$ are lengthy and not particularly illuminating, and we refer the reader to \cite{Mino:2008at,Zimmerman:2011dx,Oshita:2025qmn} for their full expressions.

Plugging Eq.~\eqref{eq:SI:near_horizon_limit} into \eqref{eq:SI:Psi4}, we obtain
\begin{align}
    [r\Psi_4 (u)]_{r\to\infty,\ell m}
    &=
    \int d\omega \int dt\,
    \hat{D}_{\ell m\omega}\tilde{Z}_{\ell m\omega}
    e^{i\omega (t-u)-im\phi(t)}
    \nonumber\\
    &\quad \times
    e^{-i (k+2i\kappa)r_*(t)} .
\end{align}
The integrand has a saddle point in the $t-\omega$ plane, determined by 
\begin{align}
    &u=t-r_*(t),\\
    &\omega=\omega_G(t),
\end{align}
where the first condition reflects the direct propagation of the emitted wave. The instantaneous complex frequency $\omega_G(t)$ is given by \cite{Oshita:2025qmn}:
\begin{align}
    \omega_G(t)=2\hat{\Omega}(t)-i\hat{\kappa}(t),
\end{align}
where
\begin{subequations}
\label{eq:real_imaginary_omega_G}
    \begin{align}
        &\hat{\Omega}(t)=\frac{\beta \Omega_H+\dot{\phi}}{1+\beta}, \\
        & \hat{\kappa}=\frac{2\beta}{1+\beta}\kappa, \label{supple:eq:kappa_hat}
    \end{align}
\end{subequations}
with $\beta=  |dr_*/dt|$ denoting the radial velocity in the tortoise coordinate, and $\dot{\phi}=d\phi/dt$ the instantaneous angular velocity. 
Here we have chosen $m=2$.

\begin{figure}
        \includegraphics[width=\linewidth]{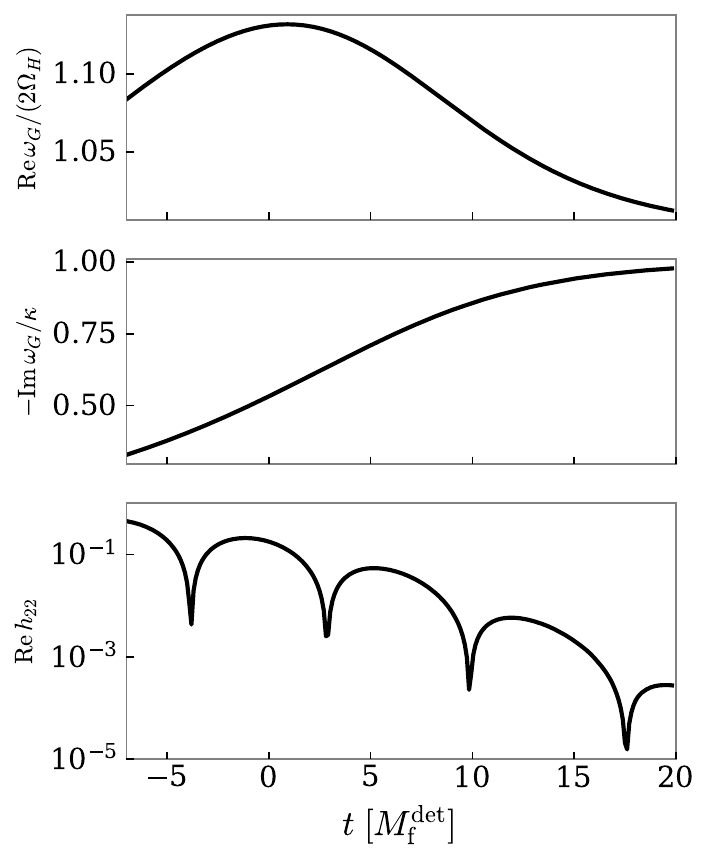}
    \caption{Time evolution of the real (top) and imaginary (middle) parts of $\omega_G(t)$, as defined in Eq.~\eqref{eq:real_imaginary_omega_G}. The resulting $\omega_G(t)$ is used to construct the direct-wave component of the GW250114-like numerical relativity waveform (bottom), corresponding to the red dashed curve in Fig.~\ref{fig:theoretical_comparison_7m}.
    }
    \label{fig:omega_G_evolution}
\end{figure}

To obtain $\omega_G(t)$ for GW250114, we evolve the geodesic equations for a test particle plunging in the equatorial plane of a Kerr black hole~\cite{o2014geometry,Mummery:2022ana}:
\begin{align}
r^2 \frac{dt}{d\tau} &= -\chi (\chi E - L_z) + \frac{r^2 + \chi^2}{\Delta} P(r)\,,\\
\frac{dr}{d\tau} &= -\sqrt{\frac{2}{3r_I}}\left(\frac{r_I}{r}-1\right)^{3/2}\,,\\
r^2 \frac{d\phi}{d\tau} &= - (\chi E-L_z) + \frac{\chi}{\Delta} P(r)\,,
\end{align}
where $P(r)= E (r^2+\chi^2) -\chi L_z$. We set $\chi = 0.672509$, the maximum-a-posteriori value inferred for the remnant black hole in GW250114~\cite{KAGRA:2025oiz,LIGOScientific:2025obp, GW250114_082203_2025}, and adopt the corresponding values of the energy $E$, angular momentum $L_z$, and radius $r_I$ of the innermost stable circular orbit:
\begin{subequations}
\begin{align}
    &E=\frac{r_I^{3/2} - 2 r_I^{1/2} + \chi}{r_I^{3/4} \sqrt{r_I^{3/2} - 3 r_I^{1/2} + 2 \chi}}, \\
    &L_z= \frac{r_I^2 - 2 \chi r_I^{1/2} + \chi^2}{r_I^{3/4}
   \sqrt{r_I^{3/2} - 3 r_I^{1/2} + 2 \chi}}.
\end{align}
\end{subequations}
The particle is initially placed at $r=r_I-10^{-5}$ and allowed to plunge into the black hole.

Figure \ref{fig:omega_G_evolution} shows the time evolution of the real (top) and imaginary (middle) parts of $\omega_G(t)$ over the same time window considered in Fig.~\ref{fig:theoretical_comparison_7m} ($t\in[-7,20]\Mfdet$), expressed in units of $2\Omega_H$ and $\kappa$, respectively. The real part of the frequency remains above $2\Omega_H$ throughout this interval, consistent with the free-frequency fits during the time interval shown in Fig.~\ref{fig:f_tau_fit}. In contrast, the imaginary part stays below $\kappa$, but grows as the particle continues to accelerate radially and $\beta$ increases, again in agreement with the trend observed in Fig.~\ref{fig:f_tau_fit}.

\begin{figure}
        \includegraphics[width=\linewidth]{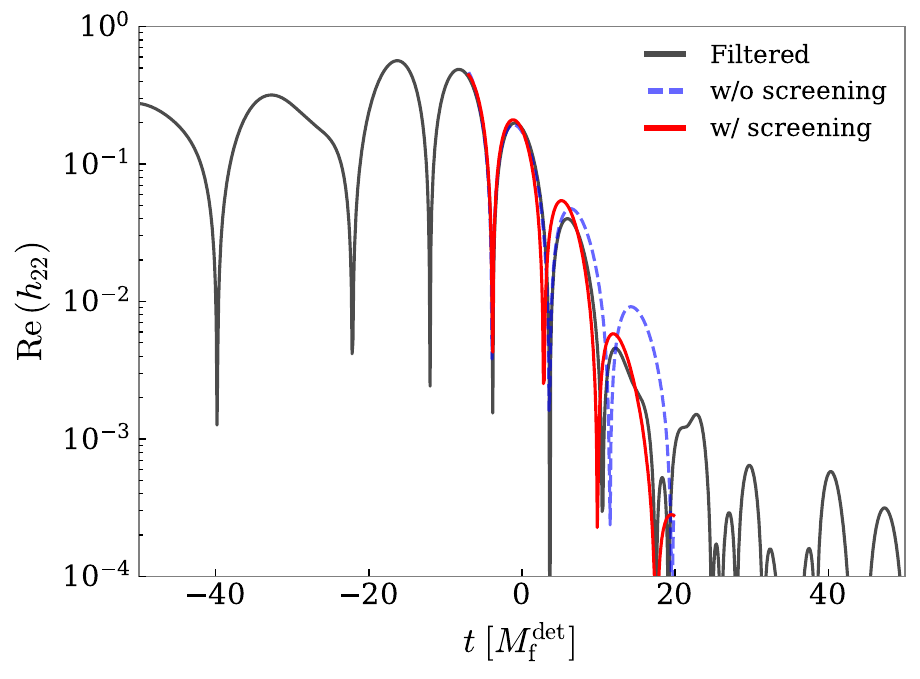}
    \caption{Comparison between the unscreened waveform model (blue dashed), $e^{-i \int \omega_G (t) dt }$, screened waveform defined in Eq.~\eqref{eq:analytic} (red solid), and the filtered quadrupolar harmonic $(\ell= m =2)$ of the NRSur7dq4 waveform (black).}
    \label{fig:SM_screening_comparison}
\end{figure}

In Fig.~\ref{fig:SM_screening_comparison}, we investigate the screening effect from the potential barrier by comparing an unscreened waveform model (blue dashed), $e^{-i \int \omega_G (t) dt }$, with the screened waveform given by Eq.~\eqref{eq:analytic}, as well as the filtered quadrupolar harmonic. The difference becomes more pronounced at late times $(t\gtrsim5\Mfdet)$, when $\omega_G(t)$ evolves sufficiently close to $\omega_H$. Within the analysis windows used in the main text, $[t_{\rm start}, t_{\rm start}+0.2\,{\rm s}]$ with $\tstart \in [-7,-3]\Mfdet$, the results are dominated by the first wave cycle. Fig.~\ref{fig:SM_screening_comparison} indicates that the screening effect plays a minor role in this early-time regime. Thus, the increasing damping rates observed in Fig.~\ref{fig:f_tau_fit} are primarily driven by the evolution of $\hat{\kappa}(t)$ [Eq.~\eqref{supple:eq:kappa_hat}], as shown in the middle panel of Fig.~\ref{fig:omega_G_evolution}.


\end{document}